\documentclass[intlimits,twoside,a4paper]{article}

\usepackage{amsmath,amssymb}
\usepackage{graphicx}
\usepackage{epsfig}

\usepackage[T2A]{fontenc}
\usepackage[cp1251]{inputenc}

\usepackage{wrapfig}

\usepackage[eqsecnum]{cmpj2}

\issue{2014}{17}{4}{43801}
\doinumber{10.5488/CMP.17.43801}

\title{Lagrangian approach in spin-oscillations problem}

\author[P.V. Pyshkin,	A.I. Kopeliovich, 	A.V. Yanovsky]
{P.V. Pyshkin,	A.I. Kopeliovich, 	A.V. Yanovsky}
\address{
B.Verkin Institute for Low Temperature Physics and Engineering \\ of the
National Academy of Sciences of Ukraine, 47 Lenin Ave., 61103 Kharkiv, Ukraine}

\date{Received June 27, 2014,  in final form July 30, 2014}
\authorcopyright{P.V. Pyshkin,	A.I. Kopeliovich, 	A.V. Yanovsky, 2014}

\begin{document}
\maketitle

\begin{abstract}
Lagrangian of electronic liquid in magneto-inhomogeneous micro-conductor has been constructed. A corresponding Euler-Lagrange equation has been solved. It was shown that the described system has eigenmodes of spin polarization and total electric current oscillations. The suggested approach permits to study the spin dynamics in an open-circuit which contains capacitance and/or inductivity.
\pacs 81.05.Xj, 75.70.Cn, 75.85.+t

\end{abstract}

At present, spintronics \cite{ref1} is one of the most active areas of investigation in solid state physics.
This interest is due, above all, to practical applications of spin-transport effects in modern microelectronics
(for example, applications of the giant magnetoresistance \cite{ref2} phenomenon). In most publications on spin
transport research, a stationary situation is assumed, for instance, the transmission of direct current through
the magnetic inhomogeneous layered conductor \cite{ref22}. But specific spin-transport effects are also possible
in a non-stationary regime. For example, in \cite{ref3}, AC spin-valve has been investigated. Another example
is the effect of the ``spin pendulum'' \cite{ref4}: the oscillations of the total current and the spin
polarization in an inhomogeneous magnetic closed conductor provided the hydrodynamic transport of electrons.
The consideration in \cite{ref4} is based on the solutions of the spin hydrodynamics equations, which in
turn were obtained using a standard Chapman-Enskog method of quasi-classical kinetic equation solution.
A description of ``spin-pendulum'' oscillations from more general principles is of independent interest
as well as it can contribute to a better understanding of spin dynamics in solid state systems. In this
article, we have developed classical
Lagrange approach to the ``spin pendulum'' oscillations of a spin electron liquid.
It allows us to study the effects of capacitance and inductivity on the ``spin pendulum'' oscillations.

\begin{figure}[!t]
\centerline{
\includegraphics[width=0.4\textwidth]{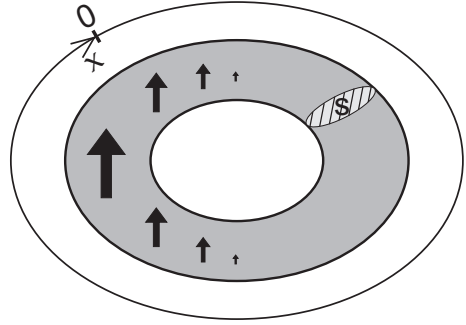}
}
\caption{Magnetically inhomogeneous closed conductor (``spin pendulum'') with a constant cross section $s$. Various sized black arrows denote spatial inhomogeneity of equilibrium spin polarization.} \label{fig1}
\end{figure}

The system which is investigated is a small-size conductor (closed or open circuit). We suppose that the length of the conductor is much bigger than the square root of its cross section and all physical values depend only on the distance $x$ along the conductor (cross section of the conductor $s$ does not depend on $x$). Let us consider a magnetically inhomogeneous conductor, when the value of magnetization of the conductor depends on $x$. That is, in terms of a spin of conducting electrons, the spin polarization of an electron density in such a conductor depends only on $x$ in the equilibrium state: $P(x)=[n_\uparrow(x) - n_\downarrow(x)]/ [n_\uparrow(x) + n_\downarrow(x)]$, where $P$ is a spin polarization and $n_{\uparrow(\downarrow)} (x)$ are equilibrium densities of electrons with spin ``up'' and ``down'' (figure~\ref{fig1}).

%
%

In particular, such a system can be realized by a spatial inhomogeneous doping with magnetic
impurities and by placing the conductor in a magnetic field. Here, we consider the magnetization
direction that is collinear along the conductor. The behavior of the conducting electrons can be
described by several macroscopic parameters: $v(x,t)$~--- the drift velocity which we consider is
common for spin-up and spin-down electrons (this is due to the hydrodynamic transport regime which
we also consider), $n_{\uparrow(\downarrow)}(x,t)=n_{0\uparrow(\downarrow)}(x)+\delta n_{\uparrow(\downarrow)} (x,t)$
are electron densities [$\delta n_{\uparrow(\downarrow)} (x,t)$ is non equilibrium perturbation of
electron densities], $\mu_{\uparrow(\downarrow)} (x,t)$ is chemical potential for spin-up and spin-down
electrons. We can express $\mu_{\uparrow(\downarrow)} (x,t)$ as a sum of the equilibrium value $\mu_0$
and perturbation $\delta\mu_{\uparrow(\downarrow)}(x,t)$ of the equilibrium chemical potential. We suppose
that all sizes of the system are  more than electronic screening length, and the characteristic frequency
is less than the plasmonic frequency. Thus, we can assume that the condition of electro-neutrality in the
conductor is: $\delta n_\uparrow (x,t) + \delta n_ \downarrow (x,t) = 0$. In the case of small
perturbation ($|\delta\mu_{\uparrow(\downarrow)}(x,t)| \ll \mu_0$), the electro-neutrality condition can be expressed as follows:
\begin{equation} \label{eq1}
\delta\mu_\uparrow N_\uparrow + \delta\mu_\downarrow N_\downarrow = 0,
\end{equation}
where $N_{\uparrow(\downarrow)}$ are equilibrium densities of states at Fermi level.

\begin{figure}[!b]
\centerline{
\includegraphics[width=0.65\textwidth]{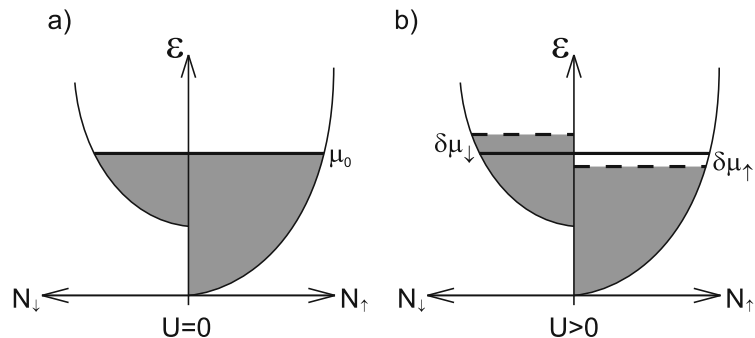}
}
\caption{Illustration of the origin of potential energy, which is connected with non-equilibrium spins. Grey filled area corresponds to the occupied electron states (at zero temperature).
} \label{fig2}
\end{figure}

Neglecting the spin-flip processes, we can write the continuity equation for electrons of each spin $\sigma$
\begin{equation}\label{eq2}
\frac{\partial \delta n_\sigma}{\partial t} + \frac{\partial (n_{0\sigma} v)}{\partial x} = 0.
\end{equation}
Let us describe the system we study by a classical Lagrange function $\mathcal{L} = T-U$ \cite{ref5}, where $T$ is the kinetic energy and $U$ is the potential energy of conducting electrons. The kinetic energy is connected with the drift of electrons:
\begin{equation*}
T = \int_V n \frac{m v^2}{2} \rd V = \frac{sj^2}{2}\int_W\frac{m}{n(x)}\rd x \, ,
\end{equation*}
where $j=nv$ is total electronic current (due to electro-neutrality $\partial j / \partial x = 0$), $m$ is an electron mass, $V$ is the total volume of the conductor, $n = n_\uparrow + n_\downarrow$, $s=\textrm{Const}$ is the cross section of the conductor, $W$ is the length of the conductor.
The potential energy $U$ is connected with the non-equilibrium spin density i.e., deviation from the equilibrium value of chemical potential~$\delta\mu_\sigma$:
\begin{equation}\label{eq4}
U = \int_V\left[ \int_{\mu_0}^{\mu_0+\delta\mu_\uparrow} \varepsilon N_{\uparrow} \rd\varepsilon + \int_{\mu_0}^{\mu_0+\delta\mu_\downarrow} \varepsilon N_{\downarrow} \rd\varepsilon \right] \rd V,
\end{equation}
where $\varepsilon$ is electron energy. In the main approximation, $N_{\uparrow(\downarrow)}$ does not depend on energy~$\varepsilon$ (but it can depend on the coordinate $x$). There is an explanatory drawing in figure~\ref{fig2} which contains schematic illustrations of equilibrium~(a) and non-equilibrium~(b) spin states which correspond to potential energy. Using electro-neutrality condition~(\ref{eq1}) and by integration~(\ref{eq4}) over energy we can transform~(\ref{eq4}) to the following expression:
\begin{equation}\label{eq5}
U = \frac{s}{2}\int \left[ N_\uparrow\delta\mu_\uparrow^2 + N_\downarrow\delta\mu_\downarrow^2 \right] \rd x.
\end{equation}
Using the fact that a drift velocity $v=j/n$ is common for two spins, we can transform~(\ref{eq2})~to:
\begin{equation}\label{eq6}
\frac{\partial \delta n_\sigma}{\partial t} + j \frac{\partial}{\partial x}\left( \frac{n_{0\sigma}}{n}\right) = 0.
\end{equation}
It is easy to integrate (\ref{eq6}) over time (note that $n_{0\sigma}$ and $n$ does not depend on time):
\begin{equation}\label{eq7}
\delta\mu_\sigma (x,t) = - \frac{1}{N_\sigma(x)}\left[ \frac{\partial}{\partial x} \frac{n_{0\sigma}(x)}{n(x)}\right] \int^t j(t') \rd t'.
\end{equation}
Based on the form of (\ref{eq7}), let us introduce a generic variable:
\begin{equation*}
 q(t) = \int^t j(t') \rd t', \qquad \dot q(t) = j(t).
\end{equation*}
Now we can write a Lagrange function in terms of $q(t)$:
\begin{equation}\label{eq9}
\mathcal{L} = \frac{s \dot q^2}{2}\int \frac{m}{n} \rd x - \frac{sq^2}{2}\int \frac{1}{N^*} \left( \frac{\rd}{\rd x}\frac{n_{0\uparrow}}{n}  \right)^2 \rd x.
\end{equation}
Using a common Lagrange equation
\begin{equation}\label{eq10}
 \frac{\rd}{\rd t}\frac{\partial \mathcal{L}}{\partial \dot q} = \frac{\partial \mathcal{L}}{\partial q}
\end{equation}
and substituting (\ref{eq9}) into (\ref{eq10}), we obtain a harmonic oscillator equation with frequency~$\omega$
\begin{equation}\label{eq11}
 \ddot q + \omega^2 q = 0, \qquad \omega^2 = \left[\int \frac{1}{N^*} \left( \frac{\rd}{\rd x}\frac{n_{0\uparrow}}{n}  \right)^2 \rd x \right] \left[\int \frac{m}{n} \rd x \right]^{-1},
\end{equation}
%
\begin{figure}
\centerline{
\includegraphics[width=0.4\textwidth]{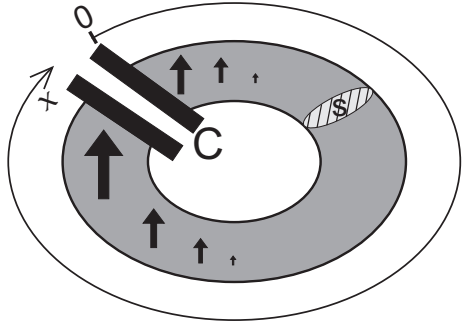}
}
\caption{Modification of ``spin pendulum''~--- magnetically inhomogeneous conductor with capacitor plates which can collect electric charge. Inductivity of the conductor is taken into account.
} \label{fig3}
\end{figure}
%
where ${N^*}^{-1} = N_\uparrow^{-1} + N_\downarrow^{-1}$. Expression (\ref{eq11}) which determines spin oscillations frequency is the same as obtained in \cite{ref4}. Notice that our approach in this work does not require the closure of the  conductor as in \cite{ref4}. Of course, a closed conductor is one of the methodologies to satisfy the electro-neutrality condition (\ref{eq1}), but now we can investigate a more general situation when we have a circuit which consists of a magnetically inhomogeneous part, capacity and inductor~(figure~\ref{fig3}). Electro-neutrality of the open-circuit is reached due to the presence of the capacity.

To take into account the presence of the capacity and inductor in the magneto-inhomogeneous conductor we must add terms into the Lagrangian.  The kinetic energy should be supplemented by a term relevant to the inductor energy~$e^2L (s\dot q)^2/2c^2$, while potential energy should be supplemented by the term relevant to the capacity energy~$e^2(sq)^2/2C$, where~$L$~and~$C$~are inductivity and capacitance of the conductor, $e$~is electron charge, $c$ is speed of light. Using~(\ref{eq10}) with the new Lagrangian we obtain the frequency of oscillations of our system:
\begin{equation}\label{eq12}
\omega^2 = \left[\int \frac{1}{N^*} \left( \frac{\rd}{\rd x}\frac{n_{0\uparrow}}{n}  \right)^2 \rd x + \frac{e^2s}{2C} \right] \left[\int \frac{m}{n} \rd x + \frac{e^2 s L}{2c^2} \right]^{-1}.
\end{equation}
It is easy to see that if we neglect the spin inhomogeneous and inertial terms, the expression~(\ref{eq12}) grades into the well known Thomson's oscillation formula for LC circuit.

As can be seen from above, we propose and describe a new device~--- an ``extended'' oscillatory circuit,
whose frequency depends on magnetic properties of the conductor.
It can be seen from the result that an external magnetic field can
be used to vary the oscillatory frequency of the device (by changing
spatial distribution of the equilibrium spin density $n_{\uparrow(\downarrow)}$).
Due to the presence of a capacity in the device, it can be easily used as a part of
external circuit by a capacitive coupling. The above described resonant properties will
appear as a peak of the impedance dependence on the frequency \cite{ref3,ref6}.
It can be seen from~(\ref{eq5}) that the potential energy is proportional to a
square of the  perturbation of a chemical potential (non-equilibrium additions
to spin densities) and to the volume~$\Delta V$ in which non-equilibrium densities
are enclosed~$U \propto \delta\mu_\sigma^2 \Delta V$. Thus, spin diffusion
(which increases~$\Delta V$, decreases $\delta\mu_\sigma$ and does not
change~$\delta\mu_\sigma \Delta V$) must be considered as a dissipative process.
Therefore, we emphasize that the expressions~(\ref{eq11}) and~(\ref{eq12})
are valid when~$\nu_\textrm{relax} \ll \omega \ll \nu_\textrm{ee},\nu_\textrm{p}$,
where~$\nu_\textrm{relax}$ is the biggest efficient frequency of relaxation processes
(frequency of collisions with a momentum loss, frequency of spin-flip processes,
characteristic diffusion frequency~$\nu_\textrm{ee}l_\textrm{ee}^2/l_\mathrm{tr}^2$,
where~$l_\textrm{ee}$ is electron free path length, $l_\mathrm{tr}$ is characteristic
length of equilibrium spatial spin inhomogeneity), $\nu_\textrm{ee}$~is the frequency
of electron-electron collisions, $\nu_\textrm{p}$~is the plasmonic frequency.

\section*{Acknowledgements}
This work was supported by the grant of the NAS of Ukraine \#4/13--N.

\vspace{-4mm}

\ukrainianpart

\title{Метод Лагранжіану в проблемі спінових осциляцій}

\author{П.В. Пишкін, О.І.Копеліович, 	А.В. Яновський}
\address{
Фізико-технічний інститут  низьких температур імені Б.І. Вєркіна НАН України,
\\ просп. Леніна, 47, 61103 Харків, Україна}

\makeukrtitle

\begin{abstract}
\tolerance=3000%
Побудовано Лагранжіан електронної рідини в магнітно-неоднорідному мікропровіднику.
Розв'язано відповідне рівняння Ейлера-Лагранжа.
Показано, що описана система має власні моди  спінової поляризації і осциляції сумарного електричного струму. Запропонований підхід
дозволяє вивчати спінову динаміку у відкритому контурі, який містить ємність і/чи індуктивність.
\keywords перенос спіну, спінтроніка

\end{abstract}

\end{document}